\documentstyle[aps,prl]{revtex}
\begin{document}
\draft
\title{ Probing Pairing Symmetry in
Sr$_2$RuO$_4$}
\author{Hae-Young Kee}
\address{
Department of Physics, University of California Los Angeles, CA 90095
}

\date{\today}
\maketitle

\begin{abstract}
We study the spin dynamics in a p-wave superconductor at the nesting
vector associated with $\alpha$ and $\beta$ bands in Sr$_2$RuO$_4$. 
We find a collective mode at the nesting vector in the superconducting
phase identified as the odd-parity pairing state which breaks
time reversal symmetry.
This mode in the spin channel only exists in the p-wave superconductor,
not in s- or d-wave superconductors.
We propose that probing this mode would clarify the pairing symmetry 
in Sr$_2$RuO$_4$.
The possibility of second superconducting phase transition is also
discussed.
%The position and the intensity of the peak is also discussed.
%The possibility of d-wave superconductor in $Sr_2RuO_4$ is also discussed.
\end{abstract}


The nature of  superconductivity  discovered in Sr$_2$RuO$_4$\cite{maeno}
has been the subject of intense theoretical and experimental
activity.
%($T_c \sim$ 1 K)\cite{maeno}.
%there has been a lot of studies to determine the nature of the 
%superconductivity in this material.
Although Sr$_2$RuO$_4$ has the same layered perovskite structure as
La$_2$CuO$_4$, the prototype of the cuprates, the behavior is remarkably
different.
At present, not much is known about any possible relation to the
cuprates.

While it is clear that electron correlation effects are important in 
Sr$_2$RuO$_4$,  
the normal state is characterized as essentially a
Fermi liquid below 50 K.
The resistivities in all directions show $T^2$ behavior for $T \leq 50 K$.
The effective mass is about $3 \sim 4 m_{electron}$ and the susceptibility
is also about $3 \sim 4 \chi_0$ where  $\chi_0$ is the pauli spin
susceptibility.
In contrast to the conventional normal state (below 50 K),
there are considerable evidences that the superconducting state (below
about 1 K) is unconventional.
The nuclear quadrupole resonance(NQR) does not show the Hebel-Slicheter peak
\cite{ishida}.
The transition temperature is very sensitive to non-magnetic
impurities.\cite{mac1}
The $^{17}$O NMR knight shift experiment shows that the spin
susceptibility has no change
across  $T_c$ but stays just the same as in the normal state
for the magnetic field parallel to the RuO$_2$ plane.\cite{ishida2}

Shortly after the discovery of the superconductivity in Sr$_2$RuO$_4$,
it was  proposed that 
the odd-pairity(spin triplet) Cooper pairs are formed in the
superconducting state in analogy with $^3$He.\cite{rice}
Further evidence favoring spin triplet pairing is the observation
of the ferromagnetic metallic state in SrRuO$_3$ which is the
three dimensional analogue of the layered Sr$_2$RuO$_4$\cite{gibb}.
Since a weak coupling analysis of the spin triplet state implies nodeless
gap\cite{rice},
it is puzzling that the specific heat and NQR measurements show a
large residual density of states (DOS), $50 \sim 60 \%$ of DOS of the
normal state, in the superconducting phase.\cite{ishida,nishizaki} 
As a consequence, nonunitary superconducting state like
$^3$He A$_1$ phase, has been proposed.\cite{sigrist1}
However, recent specific heat measurement\cite{nishizaki2}
 shows that the residual
DOS is about $25 \%$ of the normal DOS,
which indicates that the nonunitary state may not be stabilized.

An alternative explanation, so called, 
{\it orbital dependent superconductivity} 
was proposed.\cite{agt1}
Since four  $4d$ electrons in Ru$^{4+}$  partially fill the $t_{2g}$ band,
the relevant orbitals are $d_{xy}$, $d_{xz}$, and $d_{yz}$
which determine the electronic properties.
Using the quasi-two dimensional nature of the electronic dispersion,
they show that there are two superconducting order parameters
for two different classes of the orbitals.
The gap of one class of bands is substantially
smaller than that of other class of bands.
The presence of gapless excitations for temperatures greater than the smaller
gap would account for a residual DOS.  
The recent analysis of London penetration depth and coherence length
led to the evidence for orbital dependent superconductivity
identifying $d_{xy}$ as the orbital
relevant for superconductivity.\cite{riseman}
The possibility of the second superconducting phase transition
was also discussed
when the pairing symmetries are different for different classes of the
bands.

Sigrist {\it et al}  proposed\cite{sigrist2} the following  
order parameter which is claimed to be compatible with all the present
experimental data.
\begin{equation}
{\bf d} = {\hat d} (k_1 \pm i k_2),
\end{equation}
where ${\hat d}$ is parallel to the ${\hat c}$ axis and
the gap is described as the tensor represented by ${\bf d}$
in the following way. 
\begin{equation}
{\hat \Delta}({\bf k}) = \mbox{\boldmath$\sigma$} \cdot {\bf d} i \sigma_2,
\end{equation}
where \mbox{\boldmath$\sigma$} is  the Pauli matrix.
Here ${\hat d}$ is the spin vector
whose direction is perpendicular to the direction of the spin 
associated with the condensed pair.\cite{leggett}
Notice that the direction of the order parameter vector
is frozen along the ${\hat c}$ direction due to the crystal field
and  there is a full gap on the whole Fermi surface.

The details of the Fermi surface have been observed by quantum oscillations
\cite{mackenzie}.
The  Fermi surface consists of three nearly-cylindrical sheets, which
is consistent to the electronic band calculations.
\cite{oguchi}
Three Fermi sheets are labeled by $\alpha$, $\beta$, and $\gamma$.
While the $\gamma$ sheet of the Fermi surface can be attributed solely
to the $d_{xy}$ Wannier function, the $\alpha$ and $\beta$ sheets are
due to the hybridization of the $d_{xz}$ and $d_{yz}$ Wannier functions.
Combining the orbital dependent superconductivity
and experimental observation\cite{riseman},
the gap associated with $\gamma$ band is larger than that of
$\alpha$ and $\beta$ bands.
Therefore the $\gamma$ band, 
which is essentially quasi-isotropic two dimensional,
is responsible for the existing superconductivity.
On the other hand, $\alpha$ and $\beta$ sheets are quasi-one dimensional
which can be visualized as a set of parallel planes separated by
$Q=2\pi/3$ running both in $k_x$ and $k_y$ directions.
Therefore, it is natural to expect a sizable nesting effects 
at the wave vector ${\bf Q}=(2 \pi/3,2\pi/3)$ originated from
$\alpha$ and $\beta$ bands.
In the normal state, one can see that there should be a collective mode
in the spin channel due to the nesting,
and has been shown by the numerical calculation on the
static susceptibility\cite{mazin}.
The neutron scattering experiment also shows a peak
at the wave vector, $(0.6 \pi,0.6\pi,0)$ 
close to the nesting vector\cite{mazin},
with energy transfer $6.2meV$.\cite{sidis}
Mazin and Singh discussed the possibility of a competition between
p-wave and d-wave superconductivity in Sr$_2$RuO$_4$.
The experimental result\cite{sidis} also casts some doubt on the predominant
role of ferromagnetic spin fluctuations in the mechanism of spin
triplet superconductivity.
Although it is generally accepted that the pairing symmetry in Sr$_2$RuO$_4$
has the odd pairity,
a direct theoretical prediction is still necessary to determine
the pairing symmetry among the possible order parameters\cite{rice}
which have the odd pairity.

In this paper, we propose a way to probe the pairing symmetry in Sr$_2$RuO$_4$.
%assuming that all orbitals favor the same pairing symmetry as Eq. (1).
We calculate  the spin-spin correlation function at the 
nesting vector, ${\bf Q}= (2\pi/3,2\pi/3)$, using the Green function method.
It is important to  include the coupling between the spin density and 
the vectorial order parameter fluctuation 
which is the unique property of p-wave superconductor.
We find a collective mode in the spin channel in the superconducting
state only in p-wave superconductor  with the pairing symmetry
which breaks time reversal symmetry.
%which can be detected by the neutron scattering experiments\cite{neutron}.
Since the position of the resonant peak is just below $2\Delta$,
this will also determine the size of the smaller gap
related to $\alpha$ and $\beta$ bands which have the nesting. 
On the other hand,
no observation of the mode will indicate that the pairing symmetry 
associated with $\alpha$ and $\beta$ bands is different from
that with $\gamma$ band, assuming that the pairing symmetry in $\gamma$
band, which does not have any nesting effect on the Fermi surface, is
the proposed one as Eq. (1).
Therefore there must be a second superconducting phase transition
at a rather low temperature.

%One can compare the gap with the region of the temperature where 
%the residual DOS remains.
%According to the orbital dependent superconductivity, one must observe
%'no' residual DOS at low enough temperature smaller than the gap.
%The intensity of the peak depends on the strength of the
%exchange interaction, coupling constant between the spin 
%and the order parameter vector, size of the gap, and the Fermi energy.

Using the Nambu's representation, the Green function can be written as 
\cite{maki1,kee1}
\begin{equation}
G^{-1}(\omega_n,{\bf k})=i \omega_n - \xi_{{\bf k}} \rho_3 \sigma_3
-\Delta \rho_1 \mbox{\boldmath$\sigma$} \cdot {\hat d} i \sigma_2,
\end{equation}
where $\mbox{\boldmath$\rho$}$ and \mbox{\boldmath$\sigma$}
are Pauli matrices which
operate in the particle-hole and spin spaces,
respectively.
Here $\xi_{{\bf k}}= k^2/(2 m)-\mu$, where $\mu$ is the chemical
potential.
%Here we use a four-component space in which the electron-field operators
%are written in the form\cite{maki2}
%\begin{eqnarray}
%\Psi_{\bf k} &=& \left( 
%\begin{array}{cc}
%\psi_{{\bf k}\uparrow}\\ 
%\psi_{{\bf -k}\downarrow} \\
%\psi^{\dagger}_{{\bf k}\uparrow}\\ 
%\psi^{\dagger}_{{\bf -k}\downarrow} 
%\end{array}
%   \right) 
%\end{eqnarray}
%
In the superconducting state, the bare susceptibility which represents
the spin flip procedure can be written by using the Green functions.
\cite{kee1,maki2}
\begin{equation}
\chi^{00}(\omega_{\nu}, {\bf q})=T \sum_n \sum_{{\bf k}} 
Tr[G(\omega_n,{\bf k}) {\bf \alpha}_+ 
G(\omega_n+\omega_{\nu}, {\bf k}+{\bf q}) {\bf \alpha}_-],
\end{equation}
where $\omega_n$ is the Matubara frequency and 
the spin vertex \mbox{\boldmath$\alpha$} is given by\cite{kadanoff}
\begin{equation}
\mbox{\boldmath$\alpha$}= \frac{1+\rho_3}{2}\mbox{\boldmath$\sigma$}
+\frac{1-\rho_3}{2} \sigma_2 \mbox{\boldmath$\sigma$} \sigma_2,
\end{equation}
and $\alpha_{\pm}=\alpha_1 \pm i \alpha_2$.

Since the gap order parameter  fluctuation couples to the spin density,
the susceptibility renormalized by order parameter fluctuations
consists of two parts.
\begin{equation}
\chi^0(\omega,{\bf q}) = \chi^{00}(\omega,{\bf q}) -
 \frac{V(\omega,{\bf q}) g {\bar V}(\omega,{\bf q})}{1-g \Pi(\omega,{\bf q})},
\end{equation}
where the $g$ is the interaction strength and
%coupling constant between the spin density
%and the order parameter fluctuation and 
responsible for the superconductivity.
Here $V(\omega,{\bf q})$ and $\Pi(\omega,{\bf q})$ can be computed 
as follows.
\begin{eqnarray}
V (\omega_{\nu}, {\bf q}) &= &T \sum_n \sum_{\bf k} 
Tr[G(\omega_n,{\bf k}) {\bf \alpha}_+ 
G(\omega_n+\omega_{\nu}, {\bf k}+{\bf q}) ({\bf \alpha}_- \rho_1 \sigma_1)],
\nonumber\\
\Pi(\omega_{\nu}, {\bf q}) & = & T \sum_n \sum_{\bf k} 
Tr[G(\omega_n,{\bf k}) ({\bf \alpha}_+\rho_1 \sigma_1) 
G(\omega_n+\omega_{\nu}, {\bf k}+{\bf q}) ({\bf \alpha}_- \rho_1 \sigma_1)].
\end{eqnarray}

Using  $\xi_{\bf k}=-\xi_{{\bf k}+{\bf Q}}$ for
the nesting vector ${\bf Q}$, 
we found the following results at $T=0$.
\begin{eqnarray}
{\rm Re} \chi^{00} (\omega, {\bf Q}) 
&=& \cases { -g^{-1}-N_0\frac{ |\omega| \arcsin{|\omega|/2\Delta}}
{ 2\sqrt{|\omega^2-4\Delta^2|}} &
$|\omega| < 2\Delta $\cr
-g^{-1}
-N_0\frac{ |\omega| {\rm ln} 
\left(\frac{4 \Delta^2}{\omega^2-4\Delta^2} \right) }
{ 2\sqrt{\omega^2-4\Delta^2}}
& $|\omega| > 2\Delta $, \cr} 
\nonumber\\
{\rm Im} \chi^0 (\omega, {\bf Q})
&=&  \cases { 0 & $ |\omega| < 2\Delta$\cr 
-N_0\frac{\pi \omega}{2 \sqrt{\omega^2-4 \Delta^2}}
& $|\omega| > 2\Delta $, \cr}
\end{eqnarray}

\begin{eqnarray}
{\rm Re} V (\omega, {\bf Q}) 
&=& \cases { -N_0\frac{ \Delta sgn(\omega) \arcsin{|\omega|/2\Delta}}
{ \sqrt{|\omega^2-4\Delta^2|}} &
$|\omega| < 2\Delta $\cr
-N_0\frac{ \Delta sgn(\omega) {\rm ln} \left(\frac{4 \Delta^2}{\omega^2-4\Delta^2} \right) }
{ \sqrt{\omega^2-4\Delta^2}}
& $|\omega| > 2\Delta $, \cr} 
\nonumber\\
{\rm Im} V (\omega, {\bf Q})
&=&  \cases { 0 & $ |\omega| < 2\Delta$\cr 
-N_0\frac{\pi \Delta}{ \sqrt{\omega^2-4 \Delta^2}}
& $|\omega| > 2\Delta $, \cr}
\end{eqnarray}

\begin{eqnarray}
{\rm Re} \Pi (\omega, {\bf Q}) 
&=& \cases { N_0\frac{ 2\Delta^2 \arcsin{|\omega|/2\Delta}}
{|\omega| \sqrt{|\omega^2-4\Delta^2|}} &
$|\omega| < 2\Delta $\cr
N_0\frac{ 2\Delta^2 {\rm ln} \left(\frac{4 \Delta^2}{\omega^2-4\Delta^2} \right) }
{|\omega| \sqrt{\omega^2-4\Delta^2}}
& $|\omega| > 2\Delta $, \cr} 
\nonumber\\
{\rm Im} \Pi (\omega, {\bf Q})
&=&  \cases { 0 & $ |\omega| < 2\Delta$\cr 
N_0\frac{2\pi \Delta^2}{ \omega \sqrt{\omega^2-4 \Delta^2}}
& $|\omega| > 2\Delta $, \cr}
\end{eqnarray}
where $N_0$ is the DOS at the Fermi level.
Using the above result, the renormalized susceptibility is
obtained as follows for $|\omega| < 2\Delta$,
\begin{eqnarray}
{\rm Re} \chi^0(\omega, {\bf Q}) &=& -g^{-1}-\frac{N_0}{2} 
\frac{|\omega| \arcsin{\frac{|\omega|}{2\Delta}}}{\sqrt{4\Delta^2-\omega^2}}
+\frac{N_0^2 \Delta^2 |\omega| (\arcsin{\frac{|\omega|}{2\Delta}})^2}
{g^{-1} |\omega| (4\Delta^2-\omega^2)
-2 N_0 \Delta^2 \arcsin{\frac{|\omega|}{2\Delta}} \sqrt{4 \Delta^2 -\omega^2} },
\nonumber\\
{\rm Im} \chi^0(\omega, {\bf Q}) &=& 0.
\end{eqnarray}

Including the effects of the exchange interaction 
within the random phase approximation(RPA), the full 
dynamical spin susceptibility is expressed as
\begin{equation}
\chi(\omega,{\bf q})=\frac{\chi^0(\omega,{\bf q})}{1-I_{\bf q}
 \chi^0(\omega,{\bf q})},
\end{equation}
where the exchange interaction $I_{\bf Q} \equiv -I$ for
${\bf Q}$.
Since ${\rm Im} \chi^0 = 0$ and ${\rm Re} \chi^0$ diverges as 
$\omega$ approaches to $2\Delta$, there exists a  collective mode when
$I < g$.
The position of the mode is at 
\begin{equation}
\omega=2\Delta-
 \frac{\pi^2}{4} \frac{g^2 I^2}{(g-I)^2} \Delta N_0^2.
\end{equation}
Here we have assumed that the $\arcsin{|\omega|/2\Delta} \approx \pi/2$ 
consistent with the result, and
$g, I  <  1/N_0 $.
Notice that the position of the mode is very close to  $2\Delta$.
The intensity of the peak is
\begin{equation}
\frac{\pi^2}{2} \frac{g^3 I }{(g-I)^3} \Delta N_0^2.
\end{equation}

If the coupling between the spin density and the
order parameter fluctuation, $g$, is rather large\cite{note},
then there are two solutions which satisfy the condition,
Re $\chi_0 (\omega) = 1/I$, for $g < I$.
However, the separation between two modes is
$\frac{\pi^2 g^3}{4 (g-I)^2 } \sqrt{g-8 I g+8 I^2}\Delta N_0^2$ which
is very tiny so that we expect to observe only one mode at 
\begin{equation}
\omega= 2\Delta-\frac{\pi^2 g^2 (g-2 I)^2}{8 (g-I)^2}\Delta N_0^2.
\nonumber
\end{equation}

Now, let us investigate the case of the spin singlet superconductors,
such as s- or d-wave superconductor.
In the case of the spin singlet superconductor, the bare 
spin-spin correlation function can be obtained through the following
expression.
\begin{equation}
\chi^{00}(\omega,{\bf Q}) = \sum_{\bf k} \left(
1-\frac{\xi_{\bf k}\xi_{{\bf k}+{\bf Q}}
+\Delta_{\bf k} \Delta_{{\bf k}+{\bf Q}} }
{E_{\bf k} E_{{\bf k}+{\bf Q}}} \right)
\left( \frac{1}{\omega-E_{\bf k}-E_{{\bf k}+{\bf Q}}} -
\frac{1}{\omega+E_{\bf k}+E_{{\bf k}+{\bf Q}}}  \right).
\label{swave}
\end{equation}

For s-wave superconductor, i.e., $\Delta_{\bf k}=\Delta_{{\bf k}+{\bf Q}}
=\Delta$, 
one can obtain the following results for $|\omega| < 2\Delta$.
\begin{eqnarray}
{\rm Re} \chi^0(\omega, {\bf Q})& = & -N_0 
{\rm ln}(\frac{\sqrt{|\omega^2-4\Delta^2|}}{\Delta}
+\frac{\sqrt{|\omega^2-3\Delta^2|}}{\Delta} ) + {\rm ln}(C), \nonumber\\
{\rm Im} \chi^0(\omega, {\bf Q}) &=&0,
\label{sresult}
\end{eqnarray}
where $C$ is a constant.
This implies that one needs  enomously large interaction $I$ to get
the collective mode, i.e., $I \gg 1/N_0$, which is practically
impossible.

Let us study the possibility of having the resonance peak
in the d-wave superconductor at the nesting vector ${\bf Q}=(2\pi/3,2\pi/3)$.
Assuming that the superconducting phase is described by the conventional
BCS superconductor with d-wave pairing symmetry,  
$\Delta({\bf k}) =\frac{\Delta}{2} [cos(k_x)-cos(k_y)]$,  
we use the same expression as Eq. (\ref{swave}) for 
the bare  spin-spin correlation function.
%
%In high $T_c$ cuprate, it is now well known that there
%is a resonance peak at 
%${\bf Q} = (\pi, \pi,\pi)$.
%%One way of understanding this behavior is the following.
%Using the BCS superconductor with d-wave pairing symmetry,
%itight binding model for the dispersion and 
%
%We first consider the possible wave vector which produces
%the resonance peak.
Due to the coherence factor, there is a collective mode
at ${\bf Q}=(\pi,\pi)$ even without the nesting 
in the electronic dispersion.\cite{kee2}
However, in the case of ${\bf Q} = (2\pi/3, 2\pi/3)$, we do not
have simple relation as that for ${\bf Q}=(\pi,\pi)$.
In fact, $\Delta_{\bf k}$ is equal to $-\Delta_{{\bf k}+{\bf Q}}$ for
the line from ${\bf k} = (-2 \pi/3,0)$ to $(0,-2\pi/3)$, which makes
the coherence factor $O(1)$, but $\Delta_{\bf k}$ and 
$\Delta_{{\bf k}+{\bf Q}}$  have the linear dispersion in $(k_x, k_y)$ so
it does not produce any  singularity in the spin susceptibility.
If the momentum lies near the node, then we have the same dispersion relation
for the ${\bf Q}=(\pi,\pi)$ and it was found\cite{kee3} that there is 
no singularity in spin channel 
if ${\bf k}$ and ${\bf k}+{\bf Q}$ are near the nodes.
%\cite{note1}
Therefore we do not expect any collective mode in either s- or d-wave
superconductor at the wave vector ${\bf Q}=(2 \pi/3, 2\pi/3)$.

The possible Cooper pairing states 
were classified according to the irreducible representation of the tetragonal
point group D$_{4h}$ which include four one-dimensional and one two-dimensional
representations for both even and odd parity.\cite{rice,ueda} 
Assuming that the order parameter associated with $\alpha$($\beta$)
band has the odd parity with a different pairing symmetry from Eq. (1),
we also study the existence of the resonance peak with the following
order parameters classified as the odd-parity pairing.
\begin{equation}
{\bf d} = \cases { {\hat x} k_1 +{\hat y} k_2 & \cr
 {\hat x} k_1 -{\hat y} k_2  & \cr
 {\hat x} k_2 +{\hat y} k_1  & \cr
  {\hat x} k_2 -{\hat y} k_1. & \cr}
\label{odd}
\end{equation}

Due to the coherence factor, we found that the spin-spin
correlation function behaves as in Eq. \ref{sresult}.
However, once the fluctuation effect gets strong, one might get a
collective mode with the other order parameters as in Eq. \ref{odd}.
Since the coupling between the spin density and the fluctuation of 
the order parameter should not destabilize the
ground state, one needs careful investigation of the values
such as $\Delta N_0$, $g N_0$ and $I_{\bf Q} N_0$ which would determine
the possibility of the resonance peak due to the fluctuation effect.
\cite{note2}.
Therefore, we conclude that the collective mode
at the nesting vector only exits with the proposed order parameter as
in Eq. (1) unless there is unusually 
strong coupling between the spin density and
the order parameter fluctuation. 
Notice that only the order parameter which breaks the time reversal
symmetry shows the resonance peak at the nesting vector.

In conclusion, we studied the spin dynamics in p-wave superconductor
at the nesting vector associated with $\alpha$ and $\beta$ bands
in Sr$_2$RuO$_4$.
We found that there is a collective mode at the frequency just below
$2\Delta$ where $\Delta$ is a smaller gap according to the
orbital dependent superconductivity.
We show that this mode exists only in p-wave superconductor, not
in s- or d-wave superconductor.
We also presented that the other odd pairing states do not produce
the collective mode unless there is unusually strong coupling between the
spin density and the order parameter fluctuation.
Therefore we suggest that probing this mode will determine the pairing symmetry
in Sr$_2$RuO$_4$ which breaks time reversal symmetry assuming that
all bands favor the same pairing symmetry.
This will clarify the controversial situation about the possibility
of having s- or d-wave superconductor in Sr$_2$RuO$_4$.
Moreover, 
no observation of this mode will indicate that the order parameter
associated with $\alpha (\beta)$ band is different from 
the proposed order parameter as in Eq. (1). 
This implies that there must be a second superconducting phase transition
at  rather low temperature.
The observation of the strong pinning of the vortex about $50 mK$
might be an indication of second superconducting phase transition\cite{mota},
although we believe that second superconducting phase is described as one of 
the possible p-wave pairing states.
%The inelastic neutron scattering will be one of the tools to 
%measure this mode keeping  the temperature
%smaller than the gap.

\acknowledgements
I thank S. Chakravarty, Yong Baek Kim, Kazumi Maki,
 and especially M. Sigrist for helpful discussions.
This work was conducted under the auspices of the Department
of Energy, supported (in part) by funds provided by the University of
California for the conduct of discretionary research by Los Alamos
National Laboratory.


\begin{references}
\bibitem{note}  $g$ has to be bigger than about $E_F/30$.
\bibitem{note2}
Using RPA, the position of the peak in the normal state
is $\omega=\omega_c \exp{-1/(N_0 I)}$,
where $\omega_c$ is the bandwidth.
To get the observed frequency $6.2 meV$,  $1/(N_0 I)$ has to be about
$5 \sim 6$.
\end{references}
\end{document}